\documentclass{iopart}
\usepackage{iopams}
\usepackage[T1]{fontenc}
\usepackage{fourier}
\usepackage{graphicx}
\usepackage[colorlinks=true,
allcolors=blue,
dvipdfm=true]{hyperref}

\begin{document}
\title[Spin Current in Junctions composed of Multi-band SC with an SDW]{Spin Current in Junctions composed of\\
Multi-band Superconductors with a Spin-density Wave}

\author{Andreas Moor$^1$, Anatoly F~Volkov$^1$ and Konstantin B~Efetov$^{1,2}$}
\address{$^1$ Theoretische Physik III, Ruhr-Universit\"{a}t Bochum, D-44780 Bochum, Germany}
\ead{amoor@tp3.rub.de}
\address{$^2$ National University of Science and Technology ``MISiS'', Moscow, 119049, Russia}

\begin{abstract}
We calculate a nondissipative spin current and show that it can flow with or without a charge current. We consider a two-band model which can be applied to the description of Fe-based pnictides in coexistence regime of superconductivity and spin-density wave. Using quasiclassical Green's functions approach and tunneling Hamiltonian method we show that there exists a possibility to switch off the Josephson current while leaving the spin current finite. Moreover, it is possible to have the critical Josephson current and the critical spin current being proportional to each other, thus giving a possibility to measure the spin current via the Josephson current. The underlying mechanism is the interfering hopping of electron--hole pairs between different bands of the superconductors composing the junction. This is an intrinsic property of the system and provides a unique and natural way to utilize junctions made solely of pnictides in promising applications in spintronics devices.
\end{abstract}


%\keywords{spintronics, superconductor, pnictides, multiband, quasiclassical approach, tunneling Hamiltonian}

\submitto{\SUST}

\maketitle

\ioptwocol

\section{Introduction}

Control of spin currents without a net charge transfer constitutes a novel possibility to transfer and process information, and study questions of fundamental physics, and also offers a promising variety of applications~\cite{Das_Sarma_Review,Zutic_Dery_2011}. Advantages, aside from faster switching times and lower power consumption as compared to conventional devices~\cite{Sharma_2005}, are a higher integration density~\cite{Wolf_et_al_2001} and a high-bandwidth information transfer~\cite{Zutic_Dery_2011}.

A major challenge, though, are creation, detection, and control of spin currents. There are various possibilities discussed in the scientific community including the use of circularly polarized light and electrical spin injection from magnetic contacts~\cite{Das_Sarma_Review,Zutic_Dery_2011}, or creation of spin current in ferromagnet--semiconductor junctions~\cite{Ando_2011} and its enhancement in ferromagnetic insulators~\cite{Kurebayashi_2011}. Since, for practical implementations, generation and creation of spin current should not require strong magnetic fields and interfaces between semiconductors and ferromagnets, Sharma proposed a method of pure spin pumping to create spin current using a two-dimensional electron gas and a quantum cavity~\cite{Sharma_2005}. However, the detection of spin currents with arbitrary polarization direction, still constitutes a nontrivial task.

Suggestions to produce, measure, and control pure spin current in devices mostly based on semiconductors include theoretical studies of the influence of electric fields~\cite{Murakami_Nagaosa_Zhang_2003,Pareek_2004}, the optical quantum interference injection~\cite{Stevens_et_al_2003}, and mechanical torque observation~\cite{Sonin_2007}. Moreover, it has been proposed that spin current may arise in mesoscopic hybrid structures due to interfacial spin-orbit scattering~\cite{Linder_Yokoyama_2011} or at twin boundaries of noncentrosymmetric superconductors~\cite{Arahata_Neupert_Sigrist_2013}. Spin-orbit coupling is an important ingredient also in a proposal to use a single-channel voltage probe to measure spin current in lateral heterostructures~\cite{Stano_Jacquod_2011}.

Aside from coupled ferromagnets~\cite{Chen_Horsch_Manske_2014}, spin current and the Josephson effect have been investigated in triplet superconductor junctions with an insulating~\cite{Asano_2006} or ferromagnetic~\cite{Brydon_Manske_2012} layer. An interesting interplay between ferromagnetism and superconductivity influencing charge and spin current has been found in a structure consisting of two nonunitary ferromagnetic spin-triplet superconductors separated by a thin insulating layer~\cite{Gronsleth_Linder_Borven_Sudbo_2006}. More recently, spin current has been investigated in heterostructures consisting of triplet and singlet superconductors in contact with a ferromagnet~\cite{Brydon_Chen_Asano_Manske_2013}, or in a Josephson junction with double layer ferromagnets in contact with $s$-wave superconductors~\cite{Hikino_Yunoki_2013} where, by analogy to the long-range component of charge Josephson current~\cite{Bergeret_Volkov_Efetov_2001_a}, a long-range spin current may arise as a consequence of the presence of spin-triplet Cooper pairs formed by electrons of equal spin induced by the long-range proximity effect inside the ferromagnet~\cite{Bergeret_Volkov_Efetov_2001}. A similar phenomenon, i.e., the long-range penetration of the antiferromagnetic order parameter (and, correspondingly, of the spin current) into a ferromagnet, may occur in a purely magnetic structure~\cite{Moor12}.

Exceptional sources of spin currents might be provided by single-molecule magnets~\cite{Bo_et_al_2014} and topological superconductors being nowadays under intense investigation~\cite{He_et_al_2014}.

However, the discovery of superconductivity in so-called iron based pnictides~\cite{Katase_et_al_2012} revealed a novel class of high-$T_{\mathrm{c}}$ superconductors with a multiple band structure where superconductivity may coexist with spin-density wave, thus constituting a system which is intrinsically superconducting \emph{and} magnetic at the same instance. This feature has great potential for fundamental research as well as for devices based on involvement of superconducting and spin degrees of freedom.

In this Paper, we study the spin current in a Josephson junction composed of two-band superconductors with a spin-density wave separated by a thin insulating layer. We use a model which is widely accepted to describe qualitatively the coexistence region of superconductivity and spin-density wave in Fe-pnictides forming a Josephson junction~\cite{Chubukov09,Chubukov10,Schmalian10,Vavilov_Chubukov_Vorontsov_2010_SUST,Moor11}. Novel to investigations on spin current is the interplay of superconductivity and magnetism and also the possibility of tunneling of electrons and holes (and pairs of them) between nonequal bands. Despite the relative simplicity of the model, we find that, aside from the usual Josephson current, a nondissipative spin current can flow through the junction. Moreover, choosing proper conditions one can separate charge and spin degrees of freedom both intrinsic to the considered system. After the introduction of the model we provide expressions for the spin current and the corresponding expressions for the Josephson current in the case of symmetrical and asymmetrical junctions distinguished by the pairing properties of materials on each side of the junction (see next section for details). Finally, the obtained results are analyzed and an experimental setup is proposed to test them.

\section{Model}

We consider a two-band model for pnictides~\cite{Chubukov09,Chubukov10,Schmalian10,Vavilov_Chubukov_Vorontsov_2010_SUST,Moor11} suitable to qualitatively describe coexistence region of superconductivity and spin-density wave in these materials. Introducing appropriate operators~\cite{Moor11}, the Hamiltonian can be written as
\begin{equation}
    \mathcal{H} = \frac{1}{2} \sum_{\mathbf{p}} \hat{C}^{\dagger} \hat{H} \hat{C} \,,
    \label{eq:Hamiltonian}
\end{equation}
with the ${8\times 8}$~matrix~${\hat{H} = \hat{H}_{\mathrm{kin}} + \hat{H}_{\mathrm{sc}}^{\pm} + \hat{H}_{\mathrm{sdw}}}$ consisting of the corresponding parts, i.e., ${\hat{H}_{\mathrm{kin}} = \xi(\mathbf{p}) \hat{X}_{030} - \mu \hat{X}_{300}}$, ${\hat{H}_{\mathrm{sc}}^{\pm} = \xi(\mathbf{p}) \Delta^{\prime} \hat{X}_{(0,3)13} - \Delta^{\prime \prime} \hat{X}_{(3,0)23}}$, ${\hat{H}_{\mathrm{sdw}} = m \hat{X}_{113}}$, where the order parameter ${\Delta = \Delta^{\prime} + \mathrm{i} \Delta^{\prime \prime}}$ is related to the superconducting energy gap~$\Delta_1$ in the hole pocket as ${\Delta = \Delta_1^{*}}$ and the cases of $s_{++}$- and $s_{+-}$-pairing are distinguished, i.e., in the case of $s_{++}$-pairing, we have for the SC order parameter in the electron band ${\Delta_2 = \Delta_1}$, while for $s_{+-}$-pairing, ${\Delta_2 = -\Delta_1}$ (the magnitudes are taken to be equal). Moreover, we introduced the matrices~$\hat{X}_{m n \alpha}$ denoting the Kronecker product of the corresponding Pauli matrices in the band, particle--hole, and spin space, respectively,~${\hat{X}_{m n \alpha} = \rho_m \tau_n \sigma_{\alpha}}$~\cite{Moor11,Moor13a}. The dispersion relations can be linearized near the Fermi energy, ${\xi_{1,2}(\mathbf{p}) = \mp \xi(\mathbf{p}) + \mu}$ with ${\xi(\mathbf{p}) = \mathbf{v}_{\mathrm{F}} \mathbf{p}}$, the Fermi velocity~$\mathbf{v}_{\mathrm{F}}$, and the nesting parameter ${\mu = \mu_0 + \mu_{\phi} \cos (2 \phi)}$, where~$\mu_0$ describes the relative size difference of electron and hole pockets and~$\mu_{\phi}$ controls the ellipticity of the electron pocket. Moreover, the spin-density wave order parameter is defined as~${m = m_{\mathbf{q}=0} + m_{\mathbf{q}=0}^{*}}$, see~\cite{Chubukov09,Chubukov10,Schmalian10,Moor11} for further details.

In terms of the introduced operators one defines the Green's functions, e.g., the retarded and Keldysh Green's functions, and, correspondingly, the quasiclassical Green's functions for each component, ${\hat{g} = \frac{\mathrm{i}}{\pi} \int \mathrm{d} \xi \big( \hat{X}_{030} \hat{G} \big) }$.
The thus introduced quasiclassical Green's functions obey the generalized Eilenberger equation~\cite{Moor11},
\begin{equation}
    \mathbf{v}_{\mathrm{F}} \nabla \hat{g} + \big[ \omega_n \hat{X}_{030} + \mathrm{i} \hat{\Lambda} \,, \hat{g} \big] = 0 \,,
\end{equation}
supplemented with the normalization condition, ${\hat{g}^2 = 1}$. Here, the matrix~${\hat{\Lambda} = \big( -\mu \hat{X}_{330} + \hat{H}_{\mathrm{sc}}^{\pm} + \hat{H}_{\mathrm{sdw}} \big) \hat{X}_{030}}$, and ${\omega_n = (2 n + 1) \pi T}$ are the Matsubara frequencies.

The solution~${\hat{g}^{(0)} \equiv \hat{g}^{\mathrm{R}}}$ in the case of real superconducting gap and the magnetization of the spin-density wave being oriented along the $z$~axis has the form ${\hat{g}^{(0)}_{+-} = \hat{g}_{030} + \hat{g}_{100} + \hat{g}_{123} + \hat{g}_{213} + \hat{g}_{300} + \hat{g}_{323}}$ for the $s_{+-}$-pairing, and ${\hat{g}^{(0)}_{++} = \hat{\tilde{g}}_{023} + \hat{\tilde{g}}_{030} + \hat{\tilde{g}}_{123} + \hat{\tilde{g}}_{130} + \hat{\tilde{g}}_{213} + \hat{\tilde{g}}_{300}}$ for the $s_{++}$-pairing. If in contact, the phase difference of the SC condensates,~$\varphi$, and the mutual orientation of the magnetization of the SDW in the leads,~$\alpha$, can be accounted for by the functions~$\hat{g}_{\mathrm{l(r)}}$, in the left~(right) lead, respectively, which are related to~$\hat{g}^{(0)}$ via the unitary transformation ${\hat{g}_{\mathrm{l(r)}} = \hat{R}_{\pm \alpha} \hat{S}_{\pm \varphi} \hat{g}^{(0)} \hat{R}^{\dagger}_{\pm \alpha} \hat{S}^{\dagger}_{\pm \varphi}}$, where ${\hat{S}_{\pm \varphi} = \exp (\pm \mathrm{i} \hat{X}_{330} \varphi / 4)}$ and ${\hat{R}_{\pm \alpha} = \exp (\pm \mathrm{i} \hat{X}_{331} \alpha / 4)}$. They can be called the rotation matrices in Gor'kov--Nambu and spin spaces, respectively.

\begin{figure}
\centering
  \includegraphics[width=0.6\columnwidth]{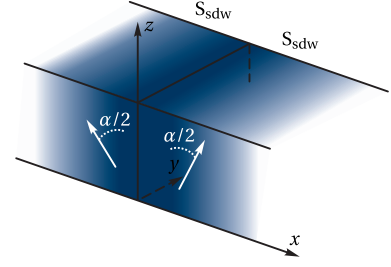}
  \caption{(Color online) Considered setup. The phases in the superconductors are assumed to be~$\pm \varphi/2$, and the angles between the $z$~axis and the magnetization vectors of the SDW are~$\pm \alpha/2$.}
  \label{fig:setup}
\end{figure}

\begin{figure}
\centering
  \includegraphics[width=1.0\columnwidth]{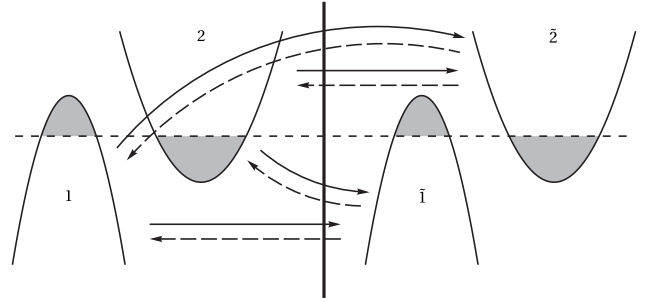}
  \caption{(Color online) Schematic tunneling through an interface between two two-band materials.}
  \label{fig:tunneling}
\end{figure}

To calculate the spin current we employ the tunneling Hamiltonian method, i.e., we consider a tunnel junction (see~\fref{fig:setup}) composed of two two-band superconductors with a spin-density wave where each SC is described by a Hamiltonian of the form~\eref{eq:Hamiltonian}, while transitions of electrons between them are accounted for by the tunneling Hamiltonian~\cite{Moor12,Moor13},
\begin{equation}
\mathcal{H}_{\mathrm{t}} = \frac{1}{2} \sum_{\mathbf{p}} \big[ \hat{C}_{\mathrm{r}}^{\dagger} \hat{\mathcal{T}} \hat{C}_{\mathrm{l}} + \mathrm{h.c.} \big] \,,
\label{eq:TunnelingHamiltonian}
\end{equation}
where the transfer matrix can be written as ${\hat{\mathcal{T}} = \hat{\mathcal{T}}_{\mathrm{qc}} \hat{X}_{030}}$ with $\hat{\mathcal{T}}_{\mathrm{qc}} = - \mathcal{T}_+ \hat{X}_{300} - \mathcal{T}_- \hat{X}_{000} + \mathrm{i} \big[ \Re(\mathcal{T}_{12}) \hat{X}_{210} + \Im(\mathcal{T}_{12}) \hat{X}_{220} \big]$, where ${\mathcal{T}_{\pm} = \big( \mathcal{T}_{11} \pm \mathcal{T}_{22} \big) / 2}$ with tunneling amplitudes~$\mathcal{T}_{ii}$ between equal bands, and~$\mathcal{T}_{12}$ describes tunneling between bands~1 and~2, cf.~\fref{fig:tunneling}. Without restriction, the tunneling elements between equal bands can be supposed as real quantities (we assume that the matrix elements~$\mathcal{T}_{ij}$ do not depend on momentum~$\mathbf{p}$), but the matrix element~$\mathcal{T}_{12}$ is, in general, complex~\cite{Moor13}.

The tunneling current in equilibrium is the nondissipative Josephson current~$I_{\mathrm{J}}$. It is given by
\begin{equation}
    I_{\mathrm{J}} = \mathrm{i} \frac{\pi e \nu_{\mathrm{l}} \nu_{\mathrm{r}}}{16} (2 \pi T) \sum_{\omega} \mathrm{Tr} \, \big\langle \hat{X}_{330} \big[\hat{\mathcal{T}}_{\mathrm{qc}} \hat{g}_{\mathrm{l}}(\omega) \hat{\mathcal{T}}_{\mathrm{qc}} \,, \hat{g}_{\mathrm{r}}(\omega)\big] \big\rangle \,,
    \label{eq:JosephsonCurrent}
\end{equation}
where~$\nu_{\mathrm{l,r}}$ and~$\hat{g}_{\mathrm{l,r}}(\omega)$ are, correspondingly, the densities of states and the quasiclassical
Green's functions in the Matsubara representation in the left and right leads, respectively, the angle brackets mean averaging over all directions of~$\mathbf{p}$, `$\mathrm{Tr}$'~denotes the operation of taking trace of a matrix, and~$e$ is the elementary charge.

In analogy to the Josephson current, the spin current can be written as
\begin{equation}
    I_{\mathrm{sp}} = \mathrm{i} \frac{\pi \mu_{\mathrm{B}} \nu_{\mathrm{l}} \nu_{\mathrm{r}}}{16} (2 \pi T) \sum_{\omega} \mathrm{Tr} \, \big\langle \hat{X}_{331} \big[\hat{\mathcal{T}}_{\mathrm{qc}} \hat{g}_{\mathrm{l}}(\omega) \hat{\mathcal{T}}_{\mathrm{qc}} \,, \hat{g}_{\mathrm{r}}(\omega)\big] \big\rangle \,,
    \label{eq:SpinCurrent}
\end{equation}
where~$\mu_{\mathrm{B}}$ is the Bohr's magneton.

\section{Spin current}

We calculate the spin current~\eref{eq:SpinCurrent} for three possible arrangements, i.e., the symmetric $s_{++} I s_{++}$, respectively, $s_{+-} I s_{+-}$ junctions, and an asymmetric $s_{++} I s_{+-}$ contact (here,~`$I$' denotes an insulating layer and~`$s$'---a superconductor with corresponding pairing symmetry in coexistence regime with a spin-density wave). The Josephson current~\eref{eq:JosephsonCurrent} has been calculated elsewhere~\cite{Moor13}, but for convenience the corresponding expressions are provided here.
\begin{enumerate}
    \item \emph{Spin current.} In both cases we obtain the Josephson-like expression,
    \begin{equation}
        I_{\mathrm{sp}} = I_{\mathrm{c}} \sin \alpha \,,
    \end{equation}
with the critical current having a different form depending on the configuration.

For the symmetric setup it reads
\begin{equation}
I_{\mathrm{c}} = I_0 + I_{\phi} \cos \phi
\end{equation}
with
\begin{eqnarray}
I_0 &\propto m_{\mathrm{l}} m_{\mathrm{r}} \big[ \mathcal{T}_{11} \mathcal{T}_{22} \pm \Re\big( \mathcal{T}_{12}^2 \big) \big] \mu_{\mathrm{B}} \nu^2 \,, \label{eq:Spin_0} \\
I_{\phi} &\propto m_{\mathrm{l}} m_{\mathrm{r}} \Delta_{\mathrm{l}} \Delta_{\mathrm{r}} \big[ \mathcal{T}_{11} \mathcal{T}_{22} \pm | \mathcal{T}_{12} |^2 \big] \mu_{\mathrm{B}} \nu^2 \,, \label{eq:Spin_phi}
\end{eqnarray}
where the upper sign denotes the $s_{++} I s_{++}$ junction and the lower---the $s_{+-} I s_{+-}$ junction.

For the asymmetric $s_{++} I s_{+-}$ setup, the critical spin current is proportional to
\begin{equation}
I_{\mathrm{c}} \propto  m_{\mathrm{l}} m_{\mathrm{r}} \big[ \mathcal{T}_{11} \mathcal{T}_{22} + \Re\big( \mathcal{T}_{12}^2 \big) \big] \mu_{\mathrm{B}} \nu^2
\end{equation}
and does not depend on the phase difference between the superconductors.

Similar expressions for the spin current were obtained in~\cite{Eremin_Nogueira_Tarento_2006} for a FFLO-like superconductor with a helimagnetic molecular field. In contrast to the assumptions of the authors of~\cite{Eremin_Nogueira_Tarento_2006}, in our case, a coexistence of superconductivity and magnetic order is intrinsic to the considered model and confirmed by experiments on Fe-based pnictides.

\item \emph{Josephson current.} In the case of a symmetric $s_{++} I s_{++}$, respectively, $s_{+-} I s_{+-}$ junction, we obtained~\cite{Moor13} the Josephson relation
    \begin{equation}
        I_{\mathrm{J}} = I_{\mathrm{c,J}} \sin \phi \,,
    \end{equation}
where
\begin{equation}
I_{\mathrm{c,J}} = I_{0,\mathrm{J}} \mp I_{\alpha,\mathrm{J}} \cos \alpha
\end{equation}
depends on the mutual orientation of the magnetization in the spin-density wave in the leads, with
\begin{eqnarray}
I_{0,\mathrm{J}} &\propto \Delta_{\mathrm{l}} \Delta_{\mathrm{r}} \big[ \mathcal{T}_{11}^2 + \mathcal{T}_{22}^2 \pm 2 \Re\big( \mathcal{T}_{12}^2 \big) \big] e \nu^2 \,, \label{eq:Jos_0J} \\
I_{\alpha,\mathrm{J}} &\propto m_{\mathrm{l}} m_{\mathrm{r}} \Delta_{\mathrm{l}} \Delta_{\mathrm{r}} \big[ \mathcal{T}_{11} \mathcal{T}_{22} \pm |\mathcal{T}_{12}|^2 \big] e \nu^2 \,. \label{eq:Jos_aJ}
\end{eqnarray}
Again, the upper sign represents the result for the $s_{++} I s_{++}$ junction and the lower---for the $s_{+-} I s_{+-}$ junction.

In the case of an asymmetric setup, an unusual so-called $\phi$-junction can be realized~\cite{Moor13},
\begin{equation}
I_{\mathrm{c,J}} = I_{1} \sin \phi + I_{2} \cos \phi \,,
\end{equation}
with ${I_1 \propto \Delta_{\mathrm{l}} \Delta_{\mathrm{r}} \big( \mathcal{T}_{11}^2 - \mathcal{T}_{22}^2 \big)}$ and ${I_2 \propto \Delta_{\mathrm{l}} \Delta_{\mathrm{r}} \Im \big( \mathcal{T}_{12}^2 \big)}$.
\end{enumerate}

\section{Analysis}

First of all, as can be seen from the definition of corresponding coefficients in the expressions for spin currents~(\ref{eq:Spin_0}) and~(\ref{eq:Spin_phi}), switching off superconductivity (in pnictides, this can be done if considering the sample at a temperature above the superconducting transition temperature but below the N\'{e}el point), we obtain a result resembling the situation considered in~\cite{Moor12} and~\cite{Chen_Horsch_Manske_2014}, i.e., a nondissipative spin current flowing through the junction due to a misalignment of the magnetization directions in the banks, where the spin current is carried by quasiparticles of same type (electrons or holes in the first case and electrons in the second case). In our model, there appears an additional component (proportiopnal to~$\mathcal{T}_{12}$) of spin (and also charge) current due to presence of two bands and due to possibility of hopping between different bands, i.e., electron--hole pairs hopping occurs with a total spin of~$1$ and no total charge transfer.

However, in coexistence regime, in the symmetric junctions, there appear corresponding contributions to the spin and charge currents due to pair hopping of Cooper pairs and Cooper-like electron--hole pairs carrying spin but no charge, leading to nontrivial mixture of charge and spin degrees of freedom. These contributions are represented by the phase dependent part~(\ref{eq:Spin_phi}) proportional to~$\Delta_{\mathrm{l}} \Delta_{\mathrm{r}}$ for the spin current, and by the angle dependent part~(\ref{eq:Jos_aJ}) proportional to~$m_{\mathrm{l}} m_{\mathrm{r}}$ for the Josephson charge current.

\begin{figure}
\centering
  \includegraphics[width=1.0\columnwidth]{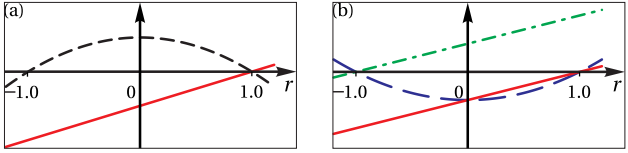}
  \caption{(Color online) Spin [solid red line denotes~$I_{\mathrm{c}}$ and~$I_0$ (respectively~$I_{\phi}$)] and Josephson [short-dashed black~($\bar{I}_{1,\mathrm{J}}$) or long-dashed blue~($I_{0,\mathrm{J}}$) and dash-dotted green~($I_{\alpha,\mathrm{J}}$) line, respectively] currents on the normalized tunneling coefficient~${r = \mathcal{T}_{22}/\mathcal{T}_{11}}$ in the asymmetric~(a) and symmetric~(b) case in arbitrary units. Other parameters are ${\Re(\mathcal{T}_{12}/\mathcal{T}_{11}) = 0}$ and ${\Im(\mathcal{T}_{12}/\mathcal{T}_{11}) = 1}$. For ${r = -1}$ the Josephson current vanishes, whereas the spin current stays finite. Note also the possibility of a sign change for the spin current as well as for the Josephson current.}
  \label{fig:currents_on_T}
\end{figure}

In this situation, the most striking result is the possibility to separate the charge and spin degrees of freedom in all considered combinations. Notably, this is an intrinsic property of the considered physical system, as opposed, e.g., to~\cite{Hikino_Yunoki_2013}, where the authors investigated a Josephson junction made of two conventional superconductors with two diffusive ferromagnetic layers sandwiched between those, by means of quasiclassical Green's functions and solving the Usadel equation for these imposing appropriate boundary conditions. In the present case, coexistence of superconductivity and magnetism is intrinsic to the considered system, but in spite of that, they can be separated, as we now show based on an analysis of corresponding coefficients of spin ($I_0$,~$I_{\phi}$ and~$I_{\mathrm{c}}$) and charge ($I_{0,\mathrm{J}}$,~$I_{\alpha,\mathrm{J}}$ and~$I_{\mathrm{c,J}}$) currents, see~\fref{fig:currents_on_T}.

In the case of an asymmetric $s_{++} I s_{+-}$ junction, the Josephson current vanishes if ${\mathcal{T}_{11}^2 - \mathcal{T}_{22}^2 = 0}$ and ${\Im \big( \mathcal{T}_{12}^2 \big) = 0}$. The spin current stays finite provided ${\mathcal{T}_{11} \mathcal{T}_{22} \neq - \Re \big( \mathcal{T}_{12} ^2\big)}$.

In the case of a symmetric $s_{++} I s_{++}$, respectively, $s_{+-} I s_{+-}$ junction, the Josephson current vanishes if ${\mathcal{T}_{11} \mathcal{T}_{22} \pm |\mathcal{T}_{12}|^2 = 0}$ and ${\mathcal{T}_{11}^2 + \mathcal{T}_{22}^2 \pm 2 \Re \big( \mathcal{T}_{12}^2 \big) = 0}$. In contrast, the spin current stays finite provided~${\Im \big( \mathcal{T}_{12} \big) \neq 0}$, since even if the $I_{\phi}$~part of the spin current vanishes due to the first condition here, the independent of~$\phi$ part stays finite.

Moreover, in the symmetric case it is possible to make the Josephson current vanish even without controlling the tunneling elements. Choosing the angle~$\alpha$ of the mutual orientation of the magnetization directions in the spin-density wave to be ${\alpha = \arccos \big( \pm I_{0,\mathrm{J}} / I_{\alpha,\mathrm{J}} \big)}$, provided ${I_{0,\mathrm{J}} \leq I_{\alpha,\mathrm{J}}}$, which is possible, e.g., if~$m$ is sufficiently large, the Josephson current vanishes, while the spin current stays finite.

Nevertheless, there exists another effect based on controlling the tunneling elements in the symmetric junctions. As is evident from the expressions of~$I_0$~(\ref{eq:Spin_0}) and~$I_{0,\mathrm{J}}$~(\ref{eq:Jos_0J}), if the relation ${\mathcal{T}_{11} = \pm \mathcal{T}_{22}}$ holds for $s_{++} I s_{++}$, respectively, $s_{+-} I s_{+-}$ junction, the spin and Josephson currents are related with each other,
\begin{eqnarray}
    I_{\mathrm{sp}} &= I_{\phi} \cos \phi \sin \alpha \,, \\
    I_{\mathrm{J}} &= I_{\alpha,\mathrm{J}} \cos \alpha \sin \phi \,,
\end{eqnarray}
via to the relation ${I_{\phi} = \frac{\mu_{\mathrm{B}}}{e} I_{\alpha,\mathrm{J}}}$ which follows from the comparison of the corresponding coefficients. This offers a possibility to measure the spin current via the Josephson critical current, i.e., measuring the value of~$I_{\mathrm{J}}$, the phase~$\phi$, and the angle~$\alpha$, one obtains~$I_{\mathrm{sp}}$ from this relation.

\section{Experiment proposal}

\begin{figure}
\centering
  \includegraphics[width=0.7\columnwidth]{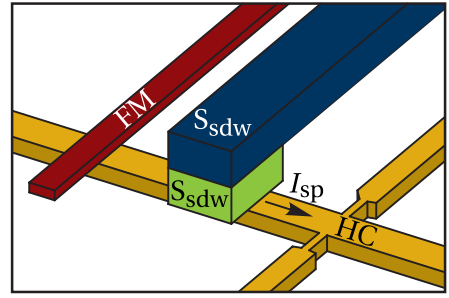}
  \caption{(Color online) Suggested experimental setting to detect the spin current in the considered Josephson junction. The Hall cross~(HC) consists of a two-band material; the red electrode~(FM) is a ferromagnet used for sample characterization~\cite{Valenzuela_Tinkham_2006}; the spin current~$I_{\mathrm{sp}}$ is injected via the contact of the considered $S_{\mathrm{sdw}}$/$S_{\mathrm{sdw}}$ junction (the two banks are shown in green and blue) with the Hall cross.}
  \label{fig:exp_setup}
\end{figure}

As the latter result is the most plausible from the experimental point of view, it may be tested by means of a nonlocal spin current measurement using a device adopted from~\cite{Valenzuela_Tinkham_2006,Kimura_Otani_Sato_Takahashi_Maekawa_2007} based on the spin Hall effect. Here, the spin current is injected replacing the ferromagnetic electrode used to inject spins in figure~1 of~\cite{Valenzuela_Tinkham_2006} by the considered Josephson junction, see~\fref{fig:exp_setup}. The used materials should be accordingly adapted, e.g., the aluminium Hall cross should be substituted with a Hall cross made of corresponding two-band material with a hole and an electron band to capture the spin current promoted by the electron-hole pairs with zero total charge.

\section{Conclusion}

We considered a two-band model for Fe-based pnictides and investigated, based on quasiclassical Green's functions approach and tunneling Hamiltonian method, Josephson junctions of two types, i.e., a symmetric configuration of superconductors with the same pairing symmetry ($s_{++} I s_{++}$ and $s_{+-} I s_{+-}$), as well as with different pairing symmetry ($s_{++} I s_{+-}$). We found that it is possible to separate the charge and spin degrees of freedom by having a finite spin current and vanishing charge current at the same time. The reverse effect is also possible but it is not discussed in details. Important ingredients are the coexistence of superconductivity and spin-density wave and the hopping between nonequal bands. Note that the possibility to obtain a superconductor with $s_{++}$~pairing symmetry out of $s_{+-}$~pairing by adding impurities has been investigated in~\cite{Efremov_Golubov_Dolgov_2013}. The discovered phenomenon may provide a reliable tool for control of spin currents, especially, as a detector or a source of spin currents based on measurement of the Josephson current, thus enabling its application in calibration of corresponding devices. Finally, we suggested an experimental setup which is suitable test the obtained results via a nonlocal spin current measurement.

\ack
We appreciate the financial support from the DFG by the Projekt~EF~11/8-1; K.~B.~E.~gratefully acknowledges the financial support of the Ministry of Education and Science of the Russian Federation in the framework of Increase Competitiveness Program of NUST~``MISiS'' (Nr.~K2-2014-015).

\section*{References}
%\bibliography{Bibliography_SUST}

\end{document}